\documentclass[aps,twocolumn,superscriptaddress]{revtex4-2}
\usepackage{amsfonts}
\usepackage{amsmath}
\usepackage{amssymb}
\usepackage{graphicx}
\usepackage{xcolor}
\usepackage{float}
\usepackage{lipsum}

\begin{document}
\title{Broadband tunable transmission non-reciprocity in thermal atoms dominated by two-photon transitions}
\author{Hui-Min Zhao}
\affiliation{Center for Quantum Sciences and School of Physics,
Northeast Normal University, Changchun 130024, China}
\author{Di-Di Zheng}
\affiliation{School of Mechanics and Optoelectronic Physics, Anhui
University of Science and Technology, Huainan 232001, China}
\author{Xiao-Jun Zhang}
\email{zhangxj037@nenu.edu.cn} \affiliation{Center for Quantum
Sciences and School of Physics, Northeast Normal University,
Changchun 130024, China}
\author{Jin-Hui Wu}
\email{jhwu@nenu.edu.cn} \affiliation{Center for Quantum Sciences
and School of Physics, Northeast Normal University, Changchun
130024, China}
\date{\today }

\begin{abstract}
We propose a scheme for realizing broadband and tunable transmission
non-reciprocity by utilizing two-photon near-resonant transitions in
thermal atoms as single-photon far-detuned transitions can be
eliminated. Our basic idea is to largely reduce the Doppler
broadenings on a pair of two-photon, probe and coupling, transitions
and meanwhile make the only four-photon transition Doppler-free
(velocity-dependent) for a forward (backward) probe field. One main
advantage of this scheme lies in that the transmission
non-reciprocity can be realized and manipulated in a frequency
range typically exceeding $200$ MHz with isolation ratio above $20$
dB and insertion loss below $1.0$ dB by modulating an assistant
field in frequency and amplitude. The intersecting angle between
four applied fields also serves as an effective control knob to
optimize the nonreciprocal transmission of a forward or backward
probe field, e.g. in a much wider frequency range approaching
$1.4$ GHz.
\end{abstract}
\maketitle

\section{Introduction}
Nonreciprocal optical
devices~\cite{APL121.261102,PRL121.203602,PRA90.043802,Nat7.579}
permitting photon transport in one direction but not in the opposite
direction, like isolators and circulators, are essential in a wide
range of modern science and technology, ranging from classical light
communications to quantum information processing. Though a lot of
advances have been made, it is still challenging to achieve optical
non-reciprocity with high isolation ratios and low insertion losses
for weak light signals due to the time-reversal symmetry of most
(linear) optical materials. Traditionally, magneto-optical media are
used to break the time-reversal symmetry with the Faraday rotation
effect, which requires however bulky magnets making against real
applications involving integrated photonic
devices~\cite{Nat5.758,PPL105.126804,PRAppl10,Optica5}. Hence,
significant efforts have been made recently to develop the
magnet-free optical non-reciprocity by exploring different
mechanisms, including nonlinear
effects~\cite{Nat9.359,Nat10,Science335,NatPh8,PRL110.234101,Nat.Electron1,PRA106.063523,Res.Phys.46,PRL.128.213605},
spatiotemporal modulations~\cite{Nat.Photon.11,Nat.Photon.3,
ACS.Photon.1,Nat.Phy.10,Nat.Photon.5}, optomechanical
interactions~\cite{Nat.Photon.10.657,Nat.568.65,Nat.Com.9.1798,Nat.Com.7.13362,Nat.Com.9.1797,Nat.Phys.13.465},
moving atomic
lattices~\cite{PRL.110.223602,PRA.92.053859,PRL.110.093901}, chiral
quantum
systems~\cite{PRL.104.163901,PRA.97.062318,Nat.541.473,PRX.5.041036,Science354},
and atomic thermal
motions~\cite{PRL125.123901,PRR2.033517,Nat.Phot12,PRL123.033902,PRAppl12.054004,PRAppl14.024032,LPR17.2022,Commun.Phys6.33,PRAppl18.024027,Sci.Adv.2021}.

Magnet-free optical isolation exploiting thermal atoms coherently
driven into the regime of electromagnetically induced transparency
(EIT) is conceptually different and admirable because relevant
realizations are simpler than those utilizing other mechanisms. In
the typical three-level $\Lambda$ configuration, for instance, the
EIT response of thermal atoms may exhibit a broken time-reversal
symmetry with the underlying quantum destructive interference
depending critically on the propagation directions of a weak probe
and a strong coupling
field~\cite{Nat.Phot12,PRL123.033902,PRAppl12.054004,PRAppl14.024032}.
That is, a forward (backward) probe field will experience a high
(low) transmissivity due to the well preserved (largely destroyed)
quantum destructive interference when its two-photon resonance along
with a forward coupling field is kept Doppler-free
(velocity-dependent), which can be operated even at the
single-photon level~\cite{Sci.Adv.2021}. Such an interesting
mechanism utilizing the direction-dependent interplay of EIT
responses and Doppler broadenings has been extended to the
four-level $N$ configuration, exhibiting a chiral cross-Kerr
nonlinear response for a probe field coming from two opposite
directions~\cite{PRL125.123901,PRL121.203602,PRR2.033517,PRL123.033902}.

Previous studies are limited to only single-photon transitions,
which can be made under appropriate conditions however negligible as
compared to two-photon transitions by eliminating some intermediate
states~\cite{PRA101.053432}. This motivates us to consider whether
the direction-dependent EIT mechanism for achieving a transmission
non-reciprocity, if extended to two-photon
transitions~\cite{PRA83.033830}, will provide some advantages and
additional degrees of manipulation? In fact, most previous
studies examine the non-reciprocal transmission of a single probe
field due to limited natural linewidths of single-photon
transitions, though it is possible to achieve the nonreciprocal
bandwidths over $100$ MHz and up to $1.0$ GHz through nonlinear
optical processes~\cite{PRL125.123901,LPR17.2022,Commun.Phys6.33}.
It is also known that the simultaneous manipulation of a vast number
of light signals is required in all-optical networks, and wavelength
division multiplexing (WDM)~\cite{WDMNatP354,WDMPR729,Appl.Sci.10}
is an effective technique for enlarging the information capacity of
optical fiber
communication~\cite{Nat.Phot12.613,PRA98.043852,OL40.2449,OC462,PRA107.053716,COL20.012701}.
Then, a specific question arises, whether the linear
non-reciprocal transmission, if extended to systems dominated by
two-photon transitions, can be realized in a wide enough frequency
range appropriate for the WDM manipulation of multiple probe
fields?

\vspace{1mm}

With above considerations, here we investigate a five-level
$\Lambda$ system for achieving two-photon EIT responses sensitive to
the propagating direction of a probe field as it can be reduced to a
three-level $\Lambda$ system with two intermediate states being
adiabatically eliminated. This is attained by making a probe and an
assistant field as well as two coupling fields to keep two-photon
near resonance, respectively, when the two pairs of oppositely
propagating fields are far detuned from relevant single-photon
transitions. Taking thermal atoms into account, we find that the
probe field incident upon one side exhibits very low losses while
that upon the other side is strongly absorbed in a frequency range
of tens of natural linewidths ($>200$ MHz) controlled by the
assistant field. This broadband transmission non-reciprocity,
facilitating WDM, of high isolation ratios ($>20$ dB) and low
insertion losses ($<1.0$ dB) is a result of the
direction-dependent Doppler effect on the only four-photon
transition and the largely reduced Doppler broadenings on both
two-photon transitions. It is also of interest that the intersecting
angle between the two pairs of oppositely traveling fields can be
tuned to realize a broader nonreciprocal bandwidth
(i.e., up to $1.4$ GHz for instance).

\section{\label{sec2}Model \& Equations}

\begin{figure*}[ptbh]
\includegraphics[width=0.7\textwidth]{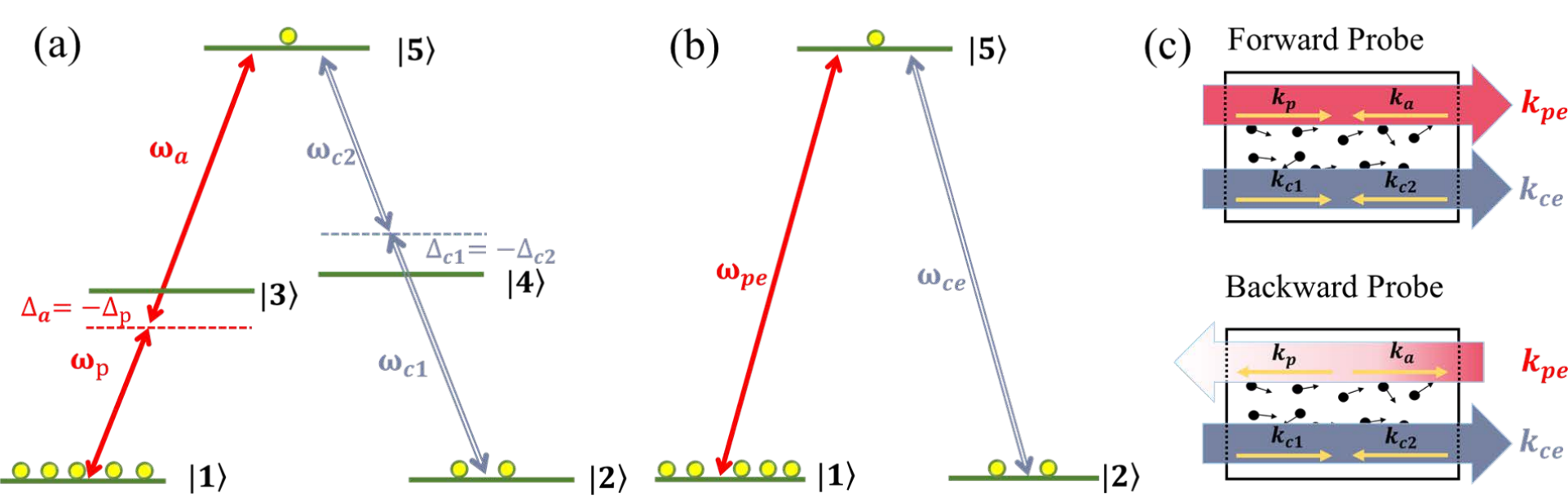}
\caption{\label{fig1}($a$) An original five-level $\Lambda$ system
driven by a probe field $\omega_{p}$, an assistant field
$\omega_{a}$, and two coupling fields $\omega_{c1,c2}$ set with
large single-photon detunings while in appropriate two-photon
resonances. ($b$) An equivalent three-level $\Lambda$ system driven
by an effective probe field $\omega_{pe}=\omega_{p}+\omega_{a}$ and
an effective coupling field $\omega_{ce}=\omega_{c1}+\omega_{c2}$ on
different two-photon transitions. ($c$) An illustration of
nonreciprocal light propagation in a warm atomic sample for
counter-traveling probe fields with wavevectors $k_{p}$ and
$-k_{p}$, respectively. Other fields have been arranged to largely
reduce Doppler broadenings on both two-photon transitions with
effective probe $\pm k_{pe}=\pm(k_{p}-k_{a})$ and coupling
$k_{ce}=k_{c1}-k_{c2}$ wavevectors.}
\end{figure*}

We consider in Fig.~\ref{fig1}($a$) a five-level atomic system
coherently driven into the $\Lambda$ configuration with two lower
ground states $|1\rangle$ and $|2\rangle$, two intermediate excited
states $|3\rangle$ and $|4\rangle$, and an upper excited state
$|5\rangle$. A probe field of frequency $\omega_{p}$ (amplitude
$E_{p}$), an assistant field of frequency $\omega_{a}$ (amplitude
$E_{a}$), a first coupling field of frequency $\omega_{c1}$
(amplitude $E_{c1}$) and a second coupling field of frequency
$\omega_{c2}$ (amplitude $E_{c2}$) act upon four transitions
$|1\rangle \leftrightarrow |3\rangle$, $|3\rangle \leftrightarrow
|5\rangle$, $|2\rangle \leftrightarrow |4\rangle$, and $|4\rangle
\leftrightarrow |5\rangle$, respectively. The corresponding (real)
Rabi frequencies are $\Omega_{p}=d_{13}E_{p}/2\hbar$,
$\Omega_{a}=d_{35}E_{a}/2\hbar$, $\Omega_{c1}=d_{24}E_{c1}/2\hbar$,
and $\Omega_{c2}=d_{45}E_{c2}/2\hbar$, respectively, with $d_{mn}$
being electric dipole moment on transition $|m\rangle
\leftrightarrow |n\rangle$. In addition, we have defined
$\Delta_{p}=\omega_{p}-\omega_{31}$ and
$\Delta_{a}=\omega_{a}-\omega_{53}$ as single-photon detunings on
the two left-arm transitions while
$\Delta_{c1}=\omega_{c1}-\omega_{42}$ and
$\Delta_{c2}=\omega_{c2}-\omega_{54}$ as single-photon detunings on
the two right-arm transitions. Then, in the electric-dipole and
rotating-wave approximations, we can write down the interaction
Hamiltonian
\begin{equation}\label{Hamilton-5}
H_{I}=-\hbar\left(
\begin{array}{ccccc}
 0 & 0 & \Omega_{p}^{*} & 0 & 0 \\
 0 & \Delta_{12} & 0 & \Omega_{c1}^{*} & 0 \\
 \Omega_{p} & 0 & \Delta_{13} & 0 & \Omega_{a}^{*} \\
 0 & \Omega_{c1} & 0 & \Delta_{14} & \Omega_{c2}^{*} \\
 0 & 0 & \Omega_{a} & \Omega_{c2} & \Delta_{15} \\
\end{array}
\right),
\end{equation}
with $\Delta_{12}=\Delta_{p}+\Delta_{a}-\Delta_{c1}-\Delta_{c2}$,
$\Delta_{13}=\Delta_{p}$,
$\Delta_{14}=\Delta_{p}+\Delta_{a}-\Delta_{c2}$, and
$\Delta_{15}=\Delta_{p}+\Delta_{a}$.

In the present work, we are only interested in the parametric
regions where single-photon detunings and Rabi frequencies are so
carefully chosen that two-photon near-resonant transitions
$|1\rangle\leftrightarrow|5\rangle$ and
$|2\rangle\leftrightarrow|5\rangle$ are dominant over single-photon
far-detuned transitions $|1\rangle\leftrightarrow|3\rangle$,
$|3\rangle\leftrightarrow|5\rangle$,
$|2\rangle\leftrightarrow|4\rangle$, and
$|4\rangle\leftrightarrow|5\rangle$. This indicates that the
five-level $\Lambda$ system in Fig.~\ref{fig1}($a$) can be reduced
to the three-level $\Lambda$ system in Fig.~\ref{fig1}($b$) under
appropriate conditions, allowing us to simplify relevant
calculations on one hand and explore new physics on the other hand.
To be more specific, in the case of $\Delta_{p}+\Delta_{a}\simeq0$,
$\Delta_{c1}+\Delta_{c2}\simeq0$, $|\Delta_{p,a}|\gg\Omega_{p,a}$,
and $|\Delta_{c1,c2}|\gg\Omega_{c1,c2}$, it is viable to derive via
the time-averaged adiabatic-elimination method~\cite{PRA83.033830}
the effective interaction Hamiltonian
\begin{equation}\label{Hamilton-3}
H_{e}=-\hbar\left(
\begin{array}{ccc}
 0 & 0 & \Omega _{pe}^* \\
 0 & \Delta _{12}+\Delta _{2d} & \Omega _{ce}^* \\
 \Omega _{pe} & \Omega _{ce} & \Delta _{15}+\Delta _{5d} \\
\end{array}
\right),
\end{equation}
where we have introduced a new Rabi frequency
$\Omega_{pe}=-\Omega_{p}\Omega_{a}/\Delta_{p}$
($\Omega_{ce}=-\Omega_{c1}\Omega_{c2}/\Delta_{c2}$) for the
effective probe (coupling) field of frequency
$\omega_{pe}=\omega_{p}+\omega_{a}$
($\omega_{ce}=\omega_{c1}+\omega_{c2}$) acting upon the two-photon
near-resonant transition $|1\rangle\leftrightarrow|5\rangle$
($|2\rangle\leftrightarrow|5\rangle$). Note also that
$\Delta_{2d}\simeq-|\Omega_{c1}|^{2}/\Delta_{c1}$
($\Delta_{5d}\simeq|\Omega_{a}|^{2}/\Delta_{a}+|\Omega_{c2}|^{2}/\Delta_{c2}$)
describes the dynamic Stark shift of state $|2\rangle$ ($|5\rangle$)
as a direct result of the virtual absorption and emission of
$\omega_{c1}$ ($\omega_{a}$ and $\omega_{c2}$) photons, which will
result in a slight deviation of the four-photon (two-photon)
resonance from $\Delta_{12}=0$ ($\Delta_{15}=0$). The validity for
eliminating states $|3\rangle$ and $|4\rangle$ can be verified by
comparing probe absorption spectra obtained by solving dynamic
equations for the five-level $\Lambda$ system starting from
Eq.~(\ref{Hamilton-5}) and those for the three-level $\Lambda$
system starting from Eq.~(\ref{Hamilton-3}) (see the
Appendix~\ref{app}).

For simplicity, here we just write down dynamic equations for the
reduced three-level $\Lambda$ system with respect to density matrix
element $\rho_{mn}$ referring to atomic population ($m=n$) or
coherence ($m\ne n$)~\cite{quantum-optics}
\begin{subequations}\label{rhomn-3}
\begin{equation}
{\partial_{t}}{\rho_{22}}=\Gamma_{52}\rho_{55}+i\Omega_{ce}^{*}\rho_{52}-i\Omega_{ce}\rho_{25},
\end{equation}
\begin{equation}
{\partial_{t}}{\rho_{11}}=\Gamma_{51}\rho_{55}+i\Omega_{pe}^{*}\rho_{51}-i\Omega_{pe}\rho_{15},
\end{equation}
\begin{equation}
{\partial_{t}}{\rho_{52}}=-g_{52}\rho_{52}+i\Omega_{pe}\rho_{12}+i\Omega_{ce}(\rho_{22}-\rho_{55}),
\end{equation}
\begin{equation}
{\partial_{t}}{\rho_{51}}=-g_{51}\rho_{51}+i\Omega_{ce}\rho_{21}+i\Omega_{pe}(\rho_{11}-\rho_{55}),
\end{equation}
\begin{equation}
{\partial_{t}}{\rho_{21}}=-g_{21}\rho_{21}+i\Omega_{ce}^{*}\rho_{51}-i\Omega_{pe}\rho_{25},
\end{equation}
\end{subequations}
which are restricted by $\rho_{ij}=\rho_{ji}^{*}$ and
$\rho_{11}+\rho_{22}+(1+\eta_{53}+\eta_{54})\rho_{55}=1$ with
$\eta_{53}\rho_{55}=\Gamma_{53}/(\Gamma_{31}+\Gamma_{32})\rho_{55}$
and
$\eta_{54}\rho_{55}=\Gamma_{54}/(\Gamma_{41}+\Gamma_{42})\rho_{55}$
accounting for populations in states $|3\rangle$ and $|4\rangle$,
respectively, due to the inevitable spontaneous decay. Above, we
have defined the complex decoherence rates
$g_{52}=\gamma_{52}+i(\Delta_{15}-\Delta_{12}+\Delta_{5d}-\Delta_{2d})$,
$g_{51}=\gamma_{51}-i(\Delta_{15}+\Delta_{5d})$ and
$g_{21}=\gamma_{21}-i(\Delta_{12}+\Delta_{2d})$ after including the
dynamic Stark shifts $\Delta_{2d}$ and $\Delta_{5d}$. Typically, the
real dephasing rate $\gamma_{mn}$ of coherence $\rho_{mn}$ depends
on the spontaneous decay rates $\Gamma_{mk}$ and $\Gamma_{nk}$ of
populations $\rho_{mm}$ and $\rho_{nn}$ through
$\gamma_{mn}=\sum_{k}(\Gamma_{mk}+\Gamma_{nk})/2$. We have also
defined $\Gamma_{51}=\Gamma_{31}\eta_{53}+\Gamma_{41}\eta_{54}$ and
$\Gamma_{52}=\Gamma_{32}\eta_{53}+\Gamma_{42}\eta_{54}$ as the
effective decay rates from state $|5\rangle$ to states $|1\rangle$
and $|2\rangle$, respectively.

In the limit of a weak effective probe field ($\Omega_{pe}\to0$),
solving Eqs.~(\ref{rhomn-3}) by setting $\partial_{t}\rho_{mn}=0$,
we can obtain the steady-state atomic population in state
$|5\rangle$
\begin{equation}\label{rho-55}
\rho_{55}=\frac{2\gamma|\Omega_{pe}|^{2}\Delta_{12e}^{2}/\Gamma}{|\Omega_{ce}|^{4}-2|\Omega_{ce}|^{2}
\Delta_{12e}\Delta_{15e}+(\Gamma^{2}+\Delta_{15e}^{2})\Delta_{12e}^{2}},
\end{equation}
with $\Delta_{12e}=\Delta_{12}+\Delta_{2d}$ and
$\Delta_{15e}=\Delta_{15}+\Delta_{5d}$ being effective detunings on
transitions $|1\rangle\leftrightarrow|2\rangle$ and
$|1\rangle\leftrightarrow|5\rangle$, respectively. We have also
considered $\Gamma_{51}=\Gamma_{52}=\Gamma$ for appropriate atomic
states and $\gamma_{51}=\gamma_{52}=\Gamma+\gamma_{l}$ with
$\gamma_{l}$ being the common linewidth of all laser
fields while neglecting the much smaller decoherence
rate $\gamma_{21}$ arising from atomic collisions. Then it is
viable to further obtain the absorption coefficient and the
probe transmissivity (see the Appendix~\ref{app}) as given below
\begin{subequations} \label{atom-cold}
\begin{equation}\label{absorp-cold}
\alpha_{p}=\frac{Nd_{13}^{2}}{\hbar\varepsilon_{0}}\frac{\pi\Gamma}{\lambda_{p}}\frac{\rho_{55}}{|\Omega_{p}|^{2}},
\end{equation}
\begin{equation}\label{transm-cold}
T_{p}=e^{-\alpha_{p}L},
\end{equation}
\end{subequations}
where $N$ ($L$) denotes the density (length) of a cold atomic sample
while $\lambda_{p}$ is the wavelength of a weak probe field
traveling through this atomic sample.

\vspace{1mm}

Taking a thermal atomic sample into account, frequencies of the
probe, assistant, and two coupling fields will be shifted to
different extents for a certain Maxwell distribution of atomic
velocity $v$ depending on their propagating directions, yielding
thus different Doppler effects. We have a total of sixteen geometric
arrangements for the four applied fields as they propagate along
either the $z$ or the $-z$ direction, two among which, referred to
as the cases of \emph{(i) forward} and \emph{(ii) backward} probes,
are of our special interest as shown in Fig.~\ref{fig1}(c). In both
cases, the probe and assistant fields as well as the two coupling
fields have been arranged to propagate in the opposite directions so
as to well suppress the Doppler broadenings on two-photon
transitions $|1\rangle\leftrightarrow|5\rangle$ and
$|2\rangle\leftrightarrow|5\rangle$. To be more specific, the
single-photon detunings should be replaced by $\Delta_{p}+k_{p}v$,
$\Delta_{a}-k_{a}v$, $\Delta_{c1}+k_{c1}v$, and
$\Delta_{c2}-k_{c2}v$ in case ($i$) while by $\Delta_{p}-k_{p}v$,
$\Delta_{a}+k_{a}v$, $\Delta_{c1}+k_{c1}v$, and
$\Delta_{c2}-k_{c2}v$ in case ($ii$) with `$+$' and `$-$' denoting
the $z$ and $-z$ directions, respectively. Here we have introduced
wavenumbers $k_{i}=2\pi/\lambda_{i}$ and wavelengths $\lambda_{i}$
for relevant light fields with $i\in\{p,a,c1,c2\}$. Accordingly, the
two-photon detunings should be replaced by $\Delta_{15}+k_{pe}v$ and
$\Delta_{15}-\Delta_{12}+k_{ce}v$ in case ($i$) while by
$\Delta_{15}-k_{pe}v$ and $\Delta_{15}-\Delta_{12}+k_{ce}v$ in case
($ii$) with largely reduced effective wavevectors
$k_{pe}=k_{p}-k_{a}$ and $k_{ce}=k_{c1}-k_{c2}$, which must result
in well suppressed Doppler broadenings on two-photon transitions. On
the other hand, the four-photon detuning should be replaced by
$\Delta_{12}+(k_{pe}-k_{ce})v$ in case ($i$) while by
$\Delta_{12}-(k_{pe}+k_{ce})v$ in case ($ii$). Hence, it is
Doppler-free and velocity-dependent, respectively, in cases ($i$)
and ($ii$) due to $k_{pe}=k_{ce}$ with $k_{p}=k_{c1}$ and
$k_{a}=k_{c2}$, which is crucial for realizing the transmission
non-reciprocity of a weak probe field.

With above considerations, it is viable to calculate the mean
populations in state $|5\rangle$ by making following integrations
for the two cases of our interest
\begin{equation}\label{rho-55-mean}
\rho_{55}^{\pm}=\int dvf(v)\rho_{55}(\pm k_{p}v,\mp
k_{a}v;k_{c1}v,-k_{c2}v),
\end{equation}
where $f(v)=e^{-v^2/v_{p}^2}/(v_{p}\sqrt{\pi})$ denotes the Maxwell
velocity distribution with $v_{p}=\sqrt{2k_{B}T/M}$ being the most
probable atomic velocity, $k_{B}$ the Boltzmann constant, $T$ the
atomic temperature, and $M$ the atomic mass. With $\rho_{55}^{\pm}$
in hand, we can further calculate the absorption coefficients and
the probe transmissivities via
\begin{subequations}\label{atom-hot}
\begin{equation}\label{absorp-hot}
\alpha_{p}^{\pm}=\frac{Nd_{13}^{2}}{\hbar\varepsilon_{0}}\frac{\pi\Gamma}{\lambda_{p}}\frac{\rho_{55}^{\pm}}{|\Omega_{p}|^{2}},
\end{equation}
\begin{equation}\label{transm-hot}
T_{p}^{\pm}=e^{-\alpha_{p}^{\pm}L},
\end{equation}
\end{subequations}
which should be direction-dependent due to
$\rho_{55}^{+}\ne\rho_{55}^{-}$. In addition to the qualitative
evaluations via $T_{p}^{+}\ne T_{p}^{-}$, the transmission
non-reciprocity for a probe field can also be quantified via two
figure of merits
\begin{subequations}
\begin{equation}\label{isolation}
\texttt{IR}=10\log_{10}\left(\frac{T_{p}^{+}}{T_{p}^{-}}\right),
\end{equation}
\begin{equation}\label{insertion}
\texttt{IL}=-10\log_{10}T_{p}^{+},
\end{equation}
\end{subequations}
referring to isolation ratio and insertion loss, respectively. Here
we have considered that $T_{p}^{+}\gg T_{p}^{-}$ around the
four-photon resonance $\Delta_{12}+\Delta_{2d}=0$ where a two-photon
EIT window can be found in case ($i$) due to the Doppler-free
arrangement but is smeared out in case ($ii$) by the residual
Doppler effect. Below, we will adopt $\texttt{IR}>20$~dB (i.e.,
$T_{p}^{+}/T_{p}^{-}>100$) and $\texttt{IL}<1.0$~dB (i.e.,
$T_{p}^{+}>0.794$) as two basic criteria for realizing a
high-performance isolator based on the transmission non-reciprocity.

It is worth noting that, effective Rabi frequencies and dynamic
Stark shifts in the reduced three-level $\Lambda$ system are also
velocity-dependent as shown below
\begin{subequations}
\begin{equation}\label{Omega-pe}
\Omega_{pe}^{\pm}=-\frac{\Omega_{p}\Omega_{a}}{\Delta_{p}\pm
k_{p}v},
\end{equation}
\begin{equation}\label{Omega-ce}
\Omega_{ce}^{\pm}=-\frac{\Omega_{c1}\Omega_{c2}}{\Delta_{c2}-
k_{c2}v},
\end{equation}
\begin{equation}\label{Delta-2d}
\Delta_{2d}^{\pm}=-\frac{|\Omega_{c1}|^{2}}{\Delta_{c1}+k_{c1}v},
\end{equation}
\begin{equation}\label{Delta-5d}
\Delta_{5d}^{\pm}=\frac{|\Omega_{a}|^{2}}{\Delta_{a}\mp
k_{a}v}+\frac{|\Omega_{c2}|^{2}}{\Delta_{c2}-k_{c2}v},
\end{equation}
\end{subequations}
which may destroy the two-photon EIT effect in case ($i$).
Fortunately, their values are very small and more importantly change
little for most atomic velocities even at a room temperature as we
choose large enough $|\Delta_{p,a,c1,c2}|$. Last but not least, the
assistant (second coupling) field may intersect the oppositely
propagating probe (first coupling) field with a misaligned angle
$180^{\circ}-\theta$ while the latter is kept to always travel along
the $\pm z$ ($z$) direction with `$+$' and `$-$' referring to cases
($i$) and ($ii$), respectively. Then, wavevectors $k_{a,c2}$ should
be replaced with the effective ones
$k_{a,c2}^{\texttt{eff}}=k_{a,c2}\cos(180^{\circ}-\theta)$ in
calculating $\rho_{55}^{\pm}$. The velocity-insensitive
$\{\Omega_{pe,ce}^{\pm},\Delta_{2d,5d}^{\pm}\}$ and angle-dependent
$k_{a,c2}^{\texttt{eff}}$ can be explored to bring additional
degrees of dynamic manipulation on the transmission non-reciprocity
discussed in the next two sections.

\begin{figure*}[ptbh]
\includegraphics[width=0.7\textwidth]{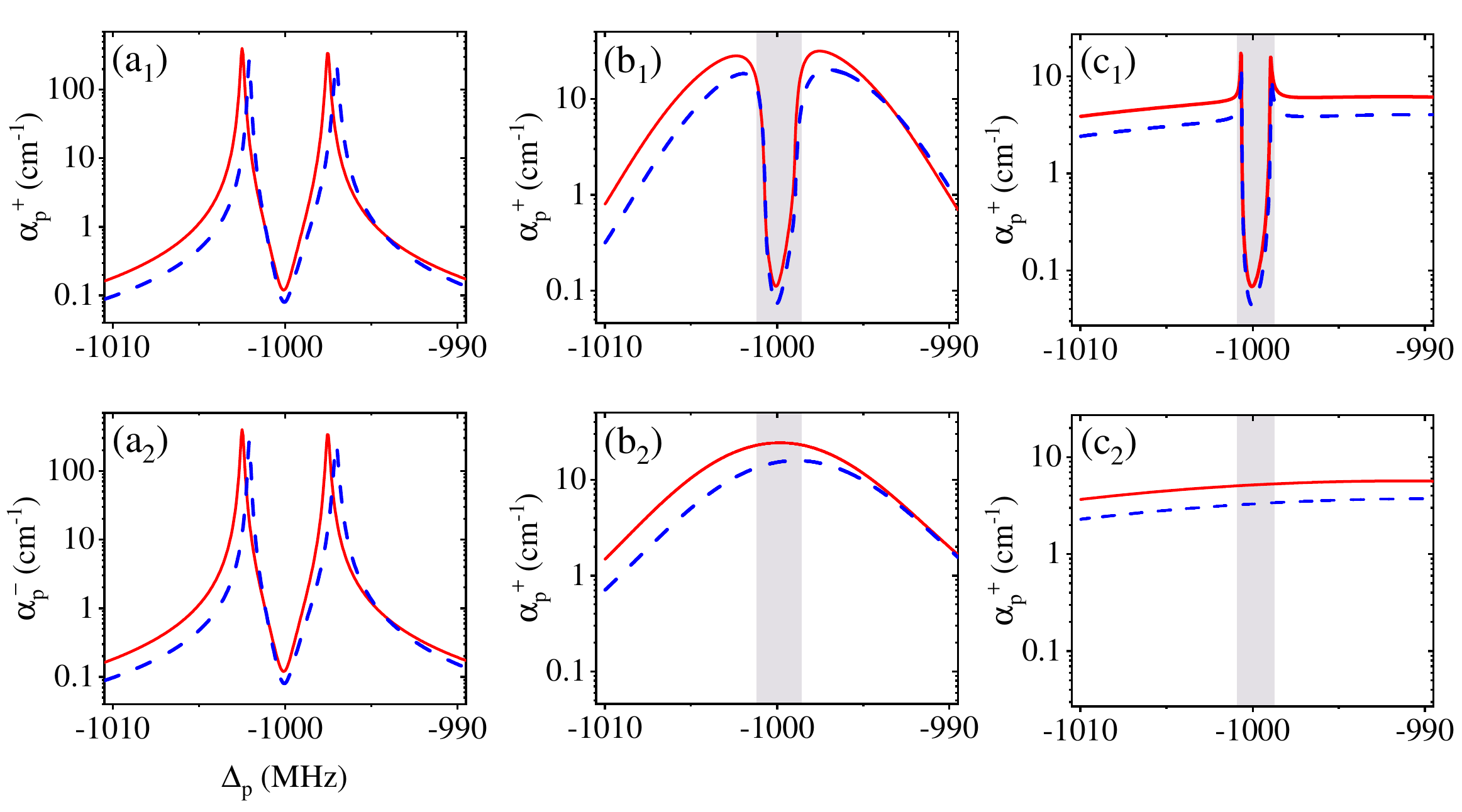}
\caption{\label{fig2} Absorption coefficients $\alpha_{p}^{+}$
(upper) and $\alpha_{p}^{-}$ (lower) as functions of
probe detuning $\Delta_{p}$ with $T=1.0$~mK (a$_{1}$,a$_{2}$),
$10$~K (b$_{1}$,b$_{2}$), and $300$~K (c$_{1}$,c$_{2}$)
for $\Omega_{a}=50$~MHz (red-solid) and $40$~MHz (blue-dashed).
Other parameters are $\Omega_{p}=0.1$~MHz,
$\Omega_{c1}=\Omega_{c2}=50$~MHz, $\Delta_{a}=\Delta_{c1}=1000$~MHz,
$\Delta_{c2}=-1002.5$~MHz, $\Gamma=0.19$~MHz,
$\gamma_{l}=0.05$~MHz, $\gamma_{21}=2.0$~kHz,
$\theta=180^{\circ}$, $N=2.0\times10^{12}$~cm$^{-3}$, $L=1.0$~cm,
$\lambda_{p}=795.0$~nm, and $d_{13}=2.537\times10^{-29}$ Cm.}
\end{figure*}

\section{\label{sec3} Broadband non-reciprocity}

In this section, we examine via numerical calculations the broadband
transmission non-reciprocity of a probe field incident upon a sample
of thermal atoms driven into the five-level $\Lambda$ configuration
in Fig.~\ref{fig1}(a) equivalent to the three-level $\Lambda$
configuration in Fig.~\ref{fig1}(b) under appropriate conditions. We
will consider $^{87}$Rb atoms as an example with states
$|1\rangle=|5S_{1/2},F=1\rangle$, $|2\rangle=|5S_{1/2},F=2\rangle$,
$|3\rangle=|5P_{1/2},F=1\rangle$, $|4\rangle=|5P_{1/2},F=2\rangle$,
and $|5\rangle=|7S_{1/2},F=2\rangle$, exhibiting transition
wavelengths $\lambda_{p}=\lambda_{c1}=795.0$ nm and
$\lambda_{a}=\lambda_{c2}=728.7$ nm as well as decay rates
$\Gamma_{31}=\Gamma_{32}=\Gamma_{41}=\Gamma_{42}=5.75$ MHz and
$\Gamma_{53}=\Gamma_{54}=0.19$ MHz~\cite{website1,website2}. 
The half Doppler broadenings can be estimated by
$\delta\omega_{D}=\sqrt{\ln2}v_{p}/\lambda_{i}$ and are about $250$
($274$) MHz on the lower (upper) single-photon transitions with
$i\in\{p,c1\}$ ($i\in\{a,c2\}$) but reduced to $22.8$ MHz on the
two-photon transitions with $i\in\{pe,ce\}$ at the temperature
$T\simeq300$ K ($v_{p}\simeq240$ m/s). It is hence appropriate to
choose $\Delta_{a,c1}/\simeq-\Delta_{p,c2}\simeq1000$~MHz and
$\Omega_{a,c1,c2}\lesssim50$~MHz so that the single-photon
transitions can be neglected for all atomic velocities while the
two-photon transitions are kept near resonances by setting
$\Delta_{15}\simeq0$ and
$\Delta_{15}-\Delta_{12}-\Delta_{12d}\simeq0$ with $\Delta_{12d}$
being small and $\Delta_{15d}$ negligible (due to
$\Delta_{a}=-\Delta_{c2}$).

\vspace{1mm}

In Fig.~\ref{fig2}, we plot two-photon absorption coefficients
$\alpha_{p}^{\pm}$ as functions of probe detuning $\Delta_{p}$ for
three typical temperatures spanning a wide range. We can see
from Figs.~\ref{fig2}(a$_{1}$,a$_{2}$) that there exist no
observable differences between $\alpha_{p}^{+}$ and $\alpha_{p}^{-}$
at a low enough temperature $T=1.0$ mK, leaving the cold atomic
sample reciprocal in absorption to a weak probe field incident from
the left or right side. Moreover, it is clear that each absorption
spectrum of $\alpha_{p}^{+}$ and $\alpha_{p}^{-}$ on the two-photon
transition $|1\rangle\leftrightarrow|5\rangle$ exhibits a typical
EIT doublet with an in-between dip at $\Delta_{p}\simeq-\Delta_{a}$,
which can be attributed to the quantum destructive interference
generated by an effective coupling field $\Omega_{ce}$ acting upon
the two-photon transition $|2\rangle\leftrightarrow|5\rangle$. As
the temperature increases to $T=10$~K in
Figs.~\ref{fig2}(b$_{1}$,b$_{2}$) or to $T=300$~K in
Figs.~\ref{fig2}(c$_{1}$,c$_{2}$), we find that
$\alpha_{p}^{+}$ and $\alpha_{p}^{-}$ become evidently different
around $\Delta_{p}\simeq-\Delta_{a}$ where the two-photon EIT dips
remain unchanged in depth, though become narrower, for
$\alpha_{p}^{+}$ but entirely disappear for $\alpha_{p}^{-}$. The
underlying physics is related to the four-photon detunings
\begin{equation}
\Delta_{12}^{\pm}=\Delta_{12}+(k_{ep}\mp k_{ec})v-\Delta_{2d}^{\pm},
\end{equation}
for $\alpha_{p}^{\pm}$ with $\Delta_{2d}^{+}=\Delta_{2d}^{-}$
depending on but insensitive to $v$. This equation indicates that
$\alpha_{p}^{+}$ is roughly Doppler-free around $\Delta_{12}=0$ due
to $k_{ep}-k_{ec}=0$ so that the EIT window remains perfect while
$\alpha_{p}^{-}$ is velocity-dependent everywhere due to
$k_{ep}+k_{ec}=2k_{ep}$ so that the EIT window entirely disappears.
It is worth noting that the increase of $\alpha_{p}^{\pm}$ with
$\Omega_{pe}$ observed for all three temperatures is a feature
absent in the three-level $\Lambda$ system dominated by
single-photon transitions and will be used later to manipulate the
transmission non-reciprocity.

\begin{figure}[ptbh]
\includegraphics[width=0.48\textwidth]{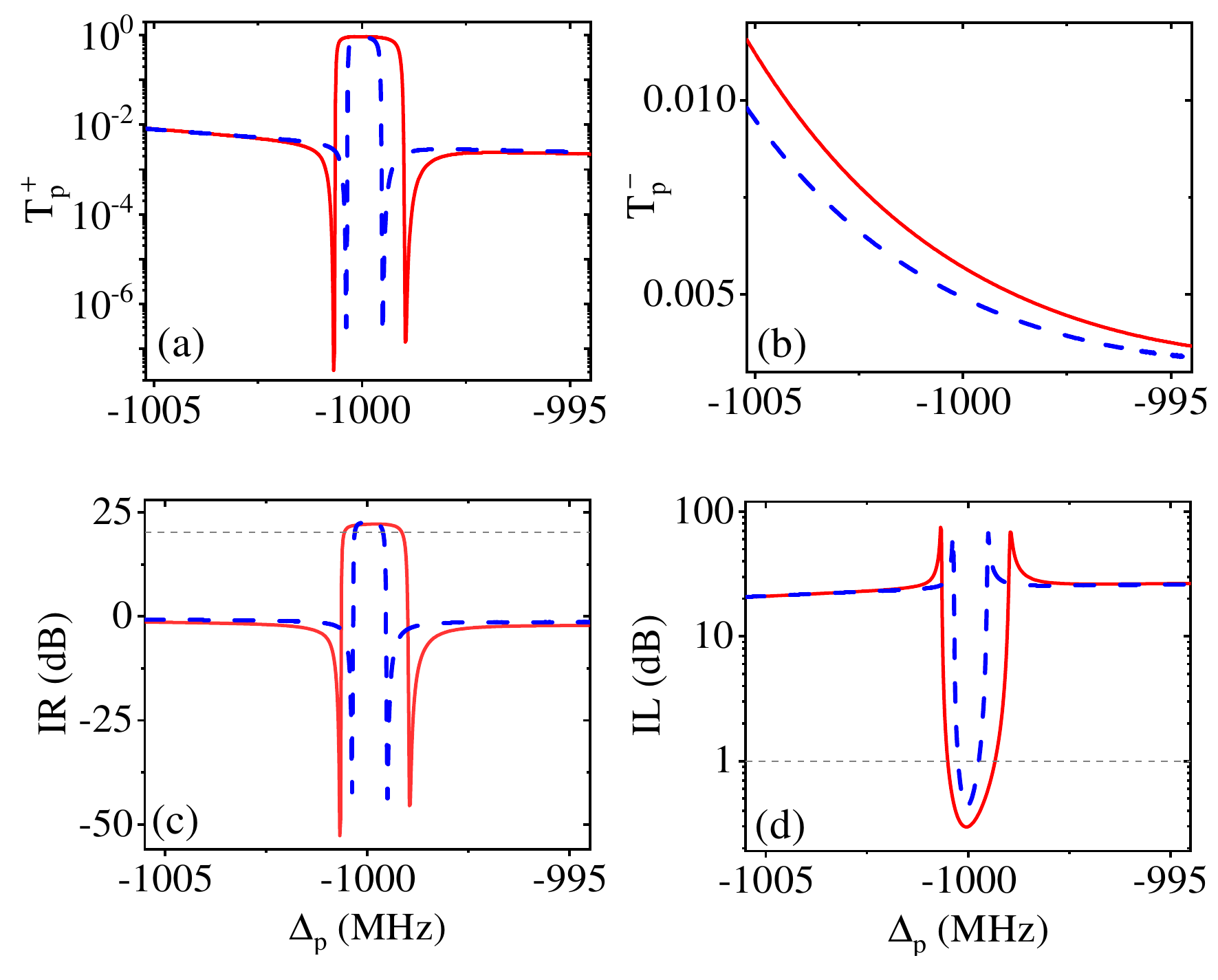}
\caption{\label{fig3} Probe transmissivities $T_{p}^{+}$ (a) and
$T_{p}^{-}$ (b) as well as isolation ratio $\texttt{IR}$ (c) and
insertion loss $\texttt{IL}$ (d) against probe detuning $\Delta_{p}$
for $\Omega_{c1}=\Omega_{c2}=50$~MHz (red-solid) and $40$~MHz
(blue-dashed). Other parameters are the same as in Fig.~\ref{fig2}
except $T=300$ K and $\Omega_{a}=50$~MHz. Gray dotted lines refer to
$\texttt{IR}=20$ dB in (c) or $\texttt{IL}=1.0$ dB in (d) as a
reference.}
\end{figure}

Above results indicate that the forward probe field can exhibit a
rather high transmissivity, while the backward probe field will be
strongly absorbed, around the four-photon resonance
$\Delta_{p}+\Delta_{a}\simeq\Delta_{c1}+\Delta_{c2}$ in a sample of
thermal atoms described by $\alpha_{p}^{\pm}$. Such an evident
transmission non-reciprocity, enabling an efficient optical
isolation, has been numerically examined in Fig.~\ref{fig3} in terms
of transmissivities $T_{p}^{\pm}$, isolation ratio $\texttt{IR}$,
and insertion loss $\texttt{IL}$ as functions of probe detuning
$\Delta_{p}$. We find from Fig.~\ref{fig3}(a) that $T_{p}^{+}$
approaches unity with a transmission bandwidth of the order of MHz
determined here by the effective coupling Rabi frequency
$\Omega_{ce}$ while Fig.~\ref{fig3}(b) shows that $T_{p}^{-}$
remains low in a much wider range though decreases slowly as
$|\Delta_{p}|$ becomes smaller due to an increase of the effective
probe Rabi frequency $\Omega_{pe}$. It is also clear from
Fig.~\ref{fig3}(c) and \ref{fig3}(d) that $\texttt{IR}$ could reach
the maximum of $22.5$ dB while $\texttt{IL}$ might be as low as
$0.3$ dB, indicating the possibility for achieving a
high-performance optical isolator. Note however that the effective
coupling Rabi frequency $\Omega_{ce}$ should not be too small,
otherwise the insertion loss will exceed $1.0$ dB on one hand and
the non-reciprocal transmission bandwidth will reduce to be
invisible on the other hand (not shown).

\begin{figure}[ptbh]
\includegraphics[width=0.46\textwidth]{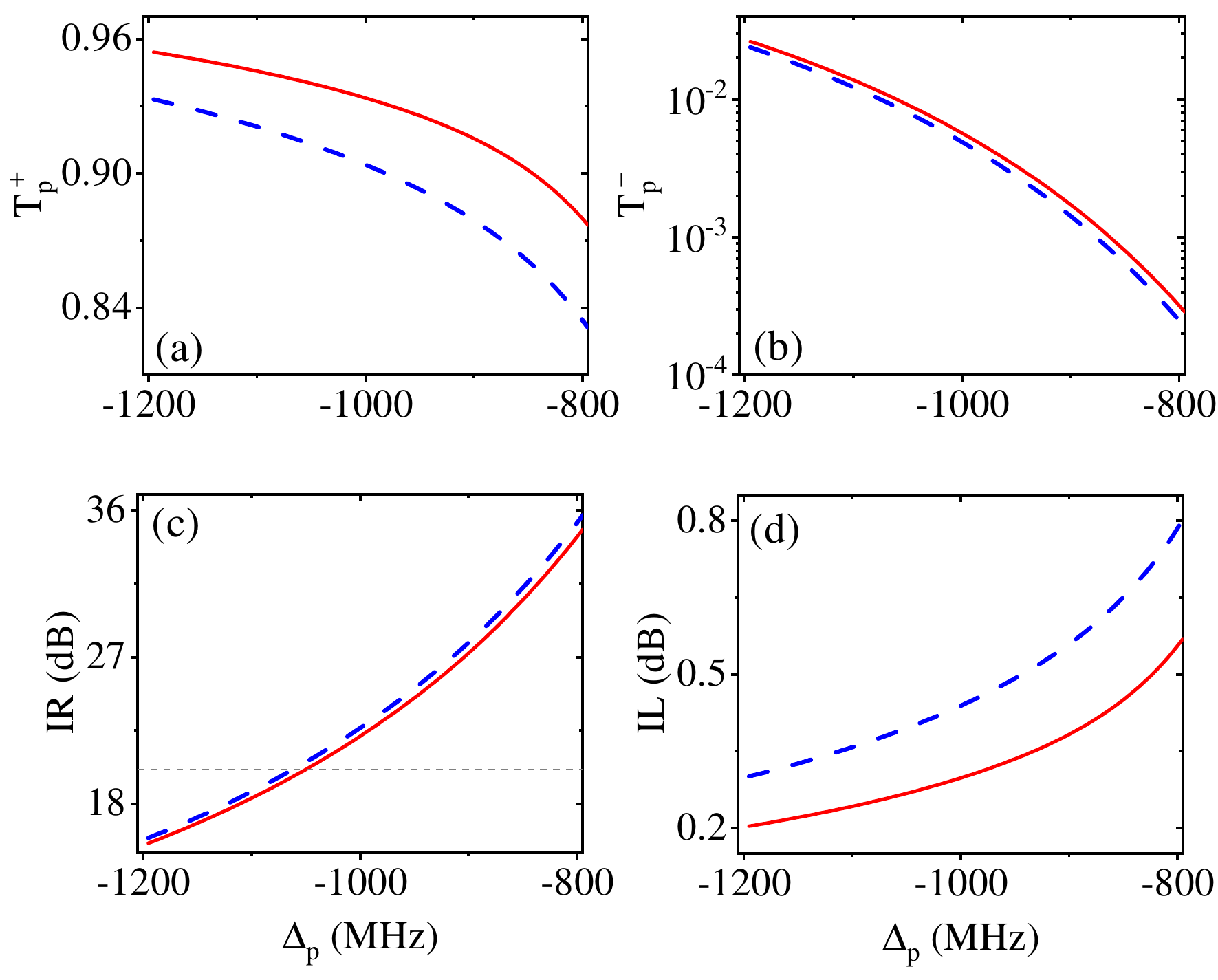}
\caption{\label{fig4} Probe transmissivities $T_{p}^{+}$ (a) and
$T_{p}^{-}$ (b) as well as isolation ratio $\texttt{IR}$ (c) and
insertion loss $\texttt{IL}$ (d) against probe detuning $\Delta_{p}$
for $\Omega_{c1}=\Omega_{c2}=50$~MHz (red-solid) and $40$~MHz
(blue-dashed). Other parameters are the same as in Fig.~\ref{fig3}
except $\Delta_{a}=-\Delta_{p}$. Gray dotted lines refer to
$\texttt{IR}=20$ dB in (c) or $\texttt{IL}=1.0$ dB in (d) as a
reference.}
\end{figure}

\begin{figure}[ptbh]
\includegraphics[width=0.46\textwidth]{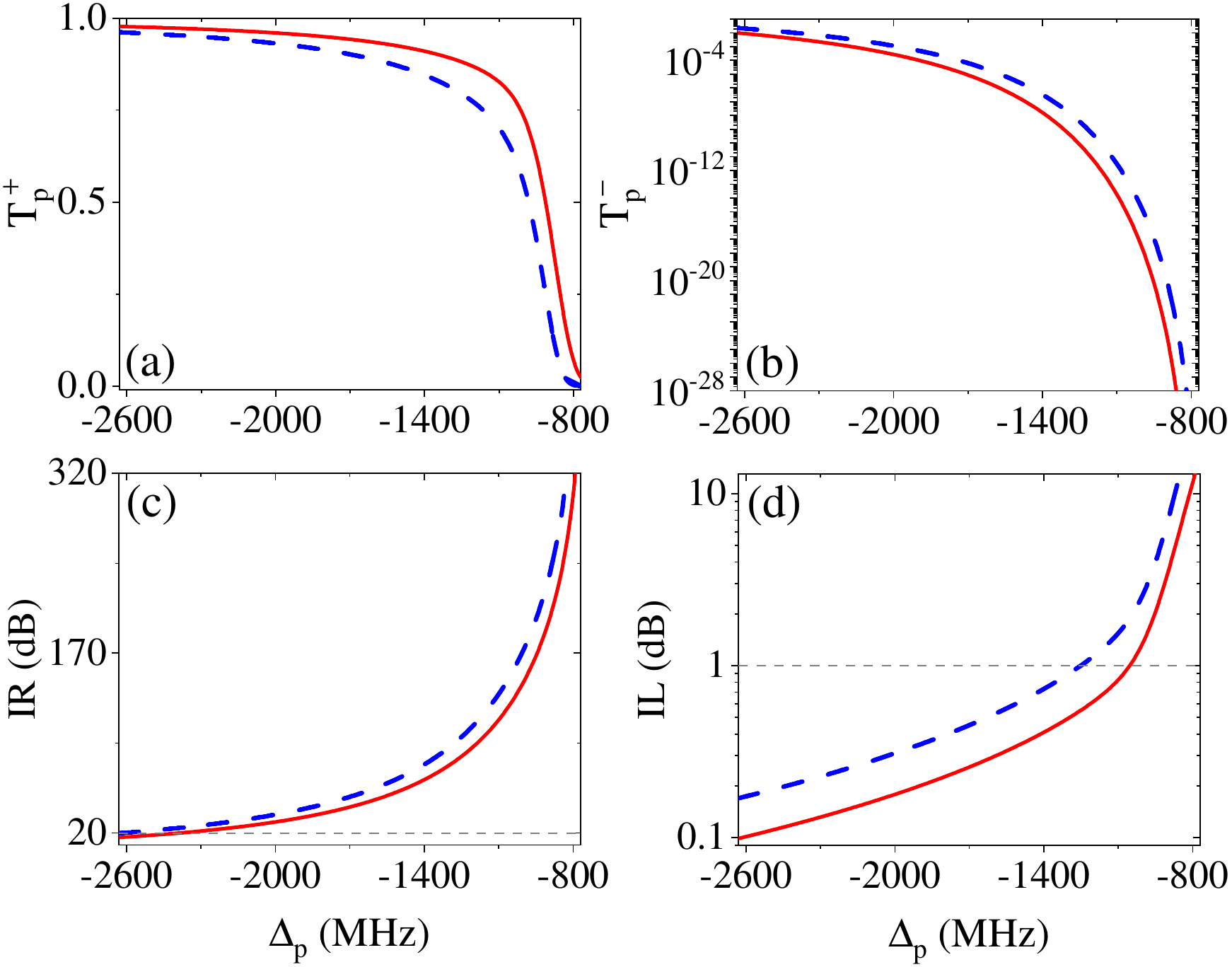}
\caption{\label{fig5} Probe transmissivities $T_{p}^{+}$ (a) and
$T_{p}^{-}$ (b) as well as isolation ratio $\texttt{IR}$ (c) and
insertion loss $\texttt{IL}$ (d) against probe detuning $\Delta_{p}$
for $\Omega_{c1}=\Omega_{c2}=50$~MHz (red-solid) and $40$~MHz
(blue-dashed). Other parameters are the same as in Fig.~\ref{fig4}
except $\theta=158^{\circ}$. Gray dotted lines refer to
$\texttt{IR}=20$ dB in (c) or $\texttt{IL}=1.0$ dB in (d) as a
reference.}
\end{figure}

We are committed in particular to achieving the nonreciprocal
transmission of a controlled broad bandwidth based on thermal atoms
dominated by two-photon transitions under consideration here.
Working in the regime of near-resonant two-photon and four-photon
transitions, the single-photon detunings can be tuned in a
relatively wide range yet without changing too much the four
important quantities plotted in Fig.~\ref{fig4}, which then
facilitates the essential WDM function in an all-optical network.
This has been examined by modulating $\Delta_{p}=-\Delta_{a}$ in the
range of $\{-1200,-800\}$~MHz to ensure that two-photon transitions
are dominant over single-photon transitions. In this range, it is
found that slight changes of effective Rabi frequency $\Omega_{pe}$
has resulted in the evident changes in $T_{p}^{\pm}$, $\texttt{IR}$,
and $\texttt{IL}$, which is impossible in a typical three-level
$\Lambda$ system. More importantly, Fig.~\ref{fig4} shows that the
isolation ratio of $\texttt{IR}>20$~dB and the insertion loss of
$\texttt{IL}<1.0$~dB can be attained in a wide frequency range of
$150\sim250$~MHz, indicating that it is viable to achieve the
nonreciprocal transmission by simultaneously handling tens of light
signals with different frequencies. Fig.~\ref{fig5} further
shows that the nonreciprocal transmission exhibits a maximal
bandwidth up to $1.4$ GHz with $\texttt{IR}>20$ and
$\texttt{IL}<1.0$~dB  as we choose $\theta=158^{\circ}$ so as to
have the smallest Doppler broadenings on both two-photon transitions
$|1\rangle\leftrightarrow|5\rangle$ and
$|2\rangle\leftrightarrow|5\rangle$. This big enlargement of
nonreciprocal bandwidth can be understood by resorting to numerical
results shown in the next section.

\section{\label{new methods} Nonreciprocal tunability}

In this section, we examine two flexible ways for manipulating the
transmission non-reciprocity by modulating additional parameters,
Rabi frequency $\Omega_{a}$ and misaligned angle $\theta$, to gain
further insights into nonreciprocal optical responses. This is
intrinsic to the reduced three-level $\Lambda$ system dominated by
two-photon near-resonant transitions and expected to facilitate the
signal or information processing in an all-optical network.

\begin{figure}[ptbh]
\includegraphics[width=0.48\textwidth]{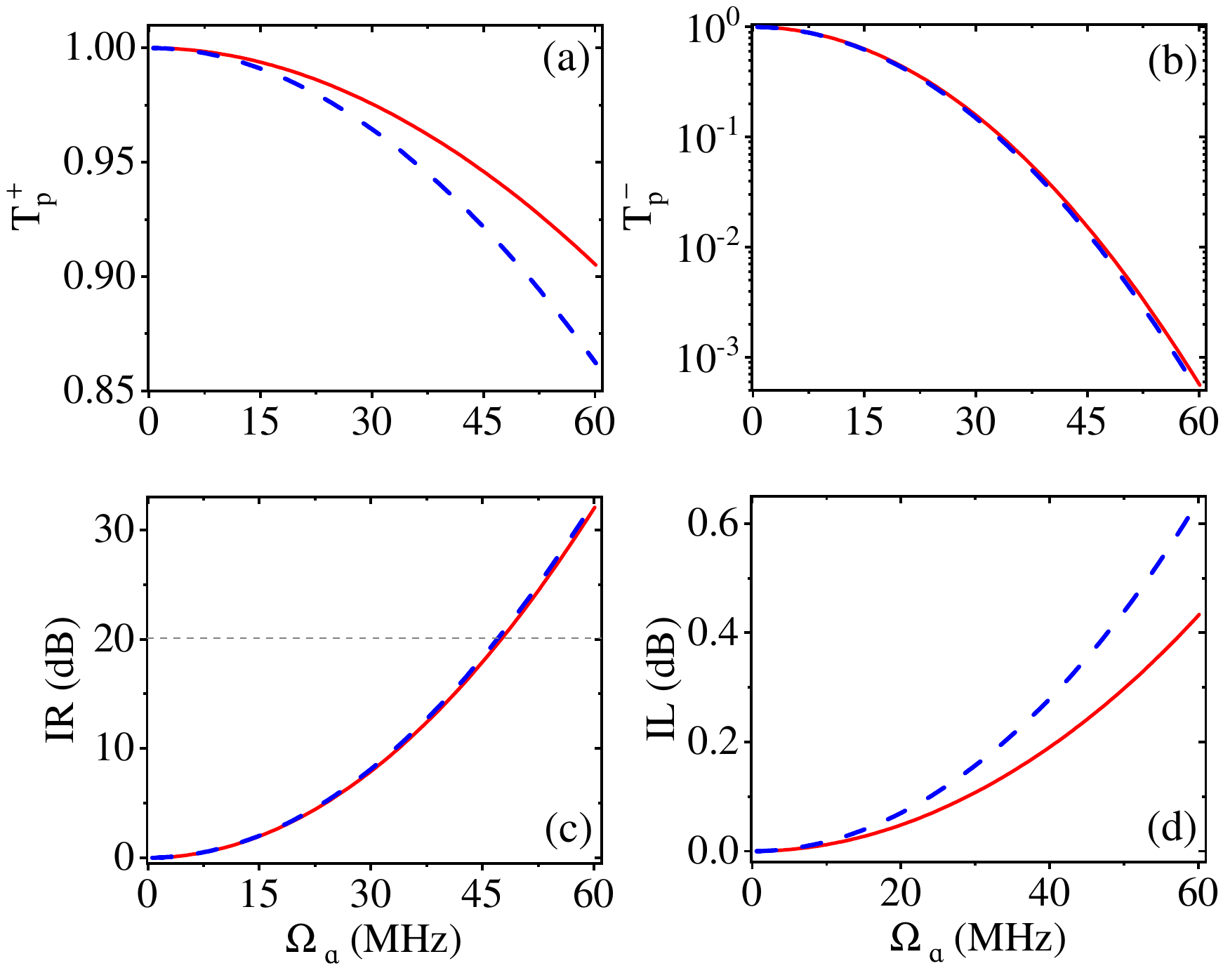}
\caption{\label{fig6} Probe transmissivities $T_{p}^{+}$ (a) and
$T_{p}^{-}$ (b) as well as isolation ratio $\texttt{IR}$ (c) and
insertion loss $\texttt{IL}$ (d) against Rabi frequency $\Omega_{a}$
for $\Omega_{c1}=\Omega_{c2}=50$~MHz (red-solid) and $40$~MHz
(blue-dashed). Other parameters are the same as in Fig.~\ref{fig3}
except $\Delta_{p}=-1000$~MHz. Gray dotted lines refer to
$\texttt{IR}=20$ dB in (c) or $\texttt{IL}=1.0$ dB in (d) as a
reference.}
\end{figure}

First, we plot in Fig.~\ref{fig6} probe transmissivities
$T_{p}^{\pm}$ together with isolation ratio $\texttt{IR}$ and
insertion loss $\texttt{IL}$ against Rabi frequency $\Omega_{a}$ of
the assistant field. It is easy to see from Figs.~\ref{fig6}(a) and
\ref{fig6}(b) that a large variation of $\Omega_{a}$ in the range of
$\{0,60\}$ MHz, corresponding to a small variation of $\Omega_{pe}$
in the range of $\{0,6\}$ kHz, will result in the evident variations
of $T_{p}^{\pm}$, which cannot be attained in a typical three-level
$\Lambda$ system independent of probe Rabi frequency $\Omega_{p}$.
We should note, however, that $T_{p}^{+}$ just reduces a few
percentage from $1.0$ to $0.90$ or $0.86$ (depending on coupling
Rabi frequencies $\Omega_{c1}=\Omega_{c2}$) while $T_{p}^{-}$
suffers a much sharper reduction from $1.0$ to $5\times10^{-4}$ or
$4\times10^{-4}$ (less sensitive to the change of
$\Omega_{c1}=\Omega_{c2}$). Accordingly, we find from
Figs.~\ref{fig6}(c) and \ref{fig6}(d) that isolation ratio
$\texttt{IR}$ increases faster than insertion loss $\texttt{IL}$
with the increase of $\Omega_{a}$, and we can achieve
$\texttt{IR}>20$~dB only with $\Omega_{a}>47.5$~MHz
($\Omega_{a}>47.0$~MHz) while $\texttt{IL}<1.0$~dB holds for
$\Omega_{a}\leq60$~MHz no matter $\Omega_{c1}=\Omega_{c2}=50$ MHz or
$\Omega_{c1}=\Omega_{c2}=40$ MHz. It is also clear that the
assistant field should be carefully modulated in order to ensure an
ideal trade-off between $\texttt{IR}$ and $\texttt{IL}$, relevant to
a high-performance optical isolator, with the working range of
$\Omega_{a}$ depending on $\Omega_{c1}=\Omega_{c2}$ for a fixed
$\Omega_{p}$.

\begin{figure}[ptbh]
\includegraphics[width=0.48\textwidth]{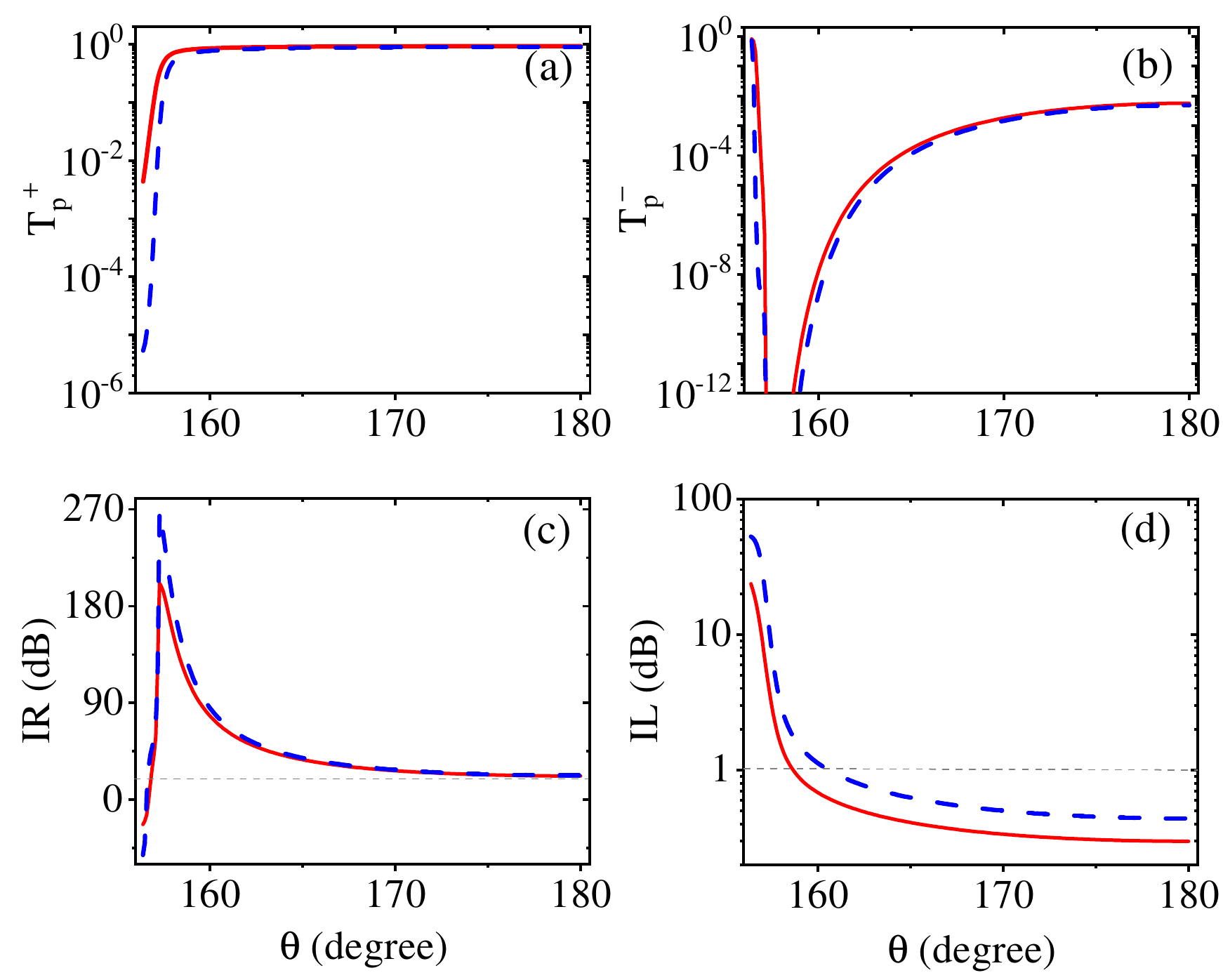}
\caption{\label{fig7} Probe transmissivities $T_{p}^{+}$ (a) and
$T_{p}^{-}$ (b) as well as isolation ratio $\texttt{IR}$ (c) and
insertion loss $\texttt{IL}$ (d) against angle $\theta$ for
$\Omega_{c1}=\Omega_{c2}=50$~MHz (red-solid) and $40$~MHz
(blue-dashed). Other parameters are the same as in Fig.~\ref{fig3}
except $\Delta_{p}=-1000$~MHz. Gray dotted lines refer to
$\texttt{IR}=20$ dB in (c) or $\texttt{IL}=1.0$ dB in (d) as a
reference.}
\end{figure}

Then, we try to plot in Fig.~\ref{fig7} probe transmissivities
$T_{p}^{\pm}$ together with isolation ratio $\texttt{IR}$ and
insertion loss $\texttt{IL}$ against misaligned angle $\theta$
between wavenumbers $k_{p}$ ($k_{c1}$) and $-k_{a}$ ($-k_{c2}$) in
the case of $\Delta_{p}=-\Delta_{a}$. Fig.~\ref{fig7}(a) shows that
$T_{p}^{+}$ decreases slowly above a quite high value until $\theta$
reduces from $180^{\circ}$ to $157.3^{\circ}$, while approaches
quickly $4\times10^{-3}$ ($5\times10^{-6}$) for
$\Omega_{c1}=\Omega_{c2}=50$ MHz ($\Omega_{c1}=\Omega_{c2}=40$ MHz)
as $\theta$ further reduces from $157.3^{\circ}$ to $156.5^{\circ}$.
Fig.~\ref{fig7}(b) shows instead that $T_{p}^{-}$ exhibits an
extremely small minimum around $\theta=157.3^{\circ}$, decreases
more evidently as $\theta$ reduces from $180^{\circ}$ to
$157.3^{\circ}$, and increases surprisingly back to $1.0$ as
$\theta$ further reduces from $157.3^{\circ}$ to $156.5^{\circ}$.
The joint variations of $T_{p}^{+}$ and $T_{p}^{-}$ due to a simple
modulation of angle $\theta$ then lead to the results shown in
Fig.~\ref{fig7}(c) and \ref{fig7}(d), where $\texttt{IR}$ exhibits a
very large maximum at $\theta=157.3^{\circ}$ while $\texttt{IL}$
increases continuously to a saturation value as $\theta$ reduces to
$156.5^{\circ}$. It is worth noting that the critical requirements
of $\texttt{IR}>20$~dB and $\texttt{IL}<1.0$~dB could be
simultaneously attained only with $\theta>158.6^{\circ}$
($\theta>160.3^{\circ}$) for $\Omega_{c1}=\Omega_{c2}=50$ MHz
($\Omega_{c1}=\Omega_{c2}=40$ MHz), while the maximum of
$\texttt{IR}$ at $\theta=157.3^{\circ}$ is meaningless as the
corresponding $\texttt{IL}$ is too large. Anyway, we can get a
better trade-off between $\texttt{IR}$ and $\texttt{IL}$ for
realizing a high-performance optical isolation by modulating
$\theta$ in the range of $\{157.3^{\circ},180^{\circ}\}$.

\begin{figure}[ptbh]
\includegraphics[width=0.48\textwidth]{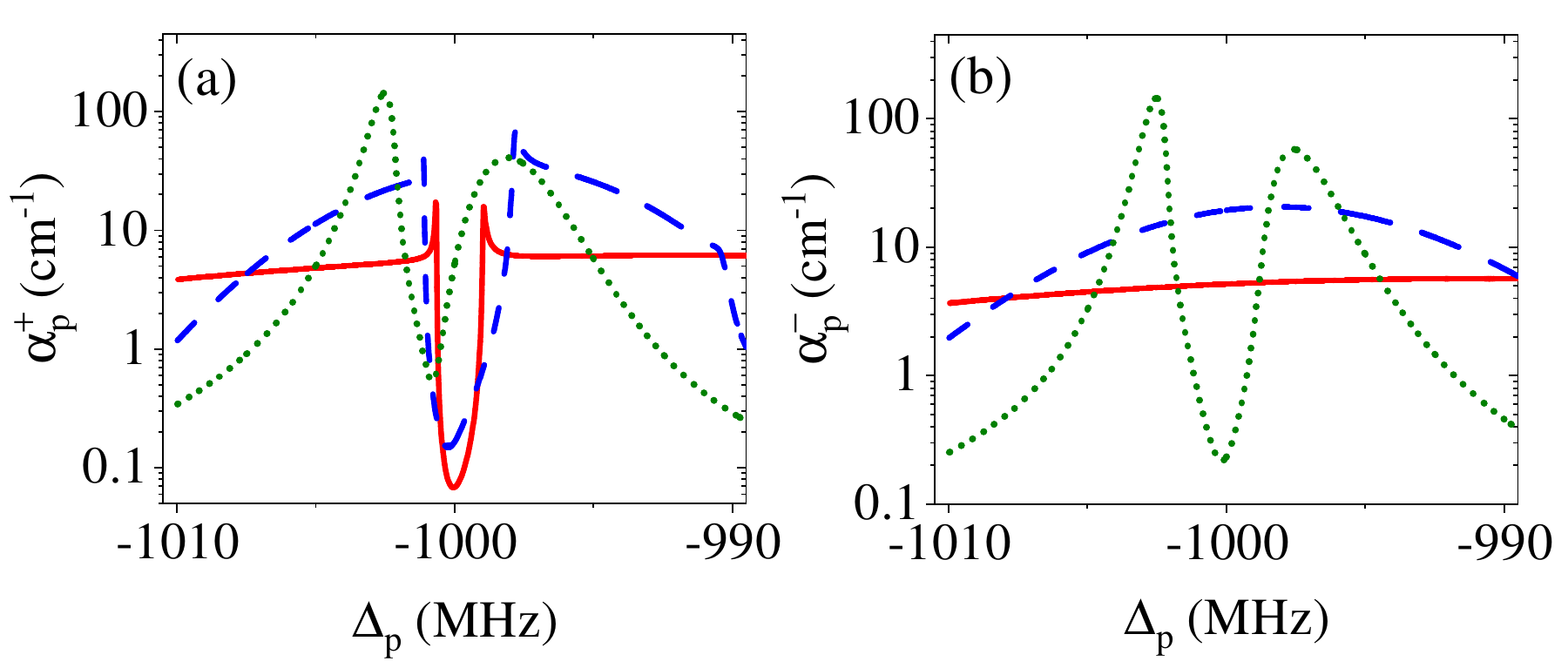}
\caption{\label{fig8} Absorption coefficients $\alpha_{p}^{+}$ (a)
and $\alpha_{p}^{-}$ (b) as functions of probe detuning $\Delta_{p}$
for $T=300$~K with $\theta=180^{\circ}$ (red-solid), $160^{\circ}$
(blue-dashed), and $156.5^{\circ}$ (green-dotted). Other parameters
are the same as in Fig.~\ref{fig3}.}
\end{figure}

To better understood what are observed in Fig.~\ref{fig7}, we
examine in Fig.~\ref{fig8} absorption coefficients
$\alpha_{p}^{\pm}$ as functions of probe detuning $\Delta_{p}$ for
three typical values of angle $\theta$. Fig.~\ref{fig8}(a) shows
that the EIT window of $\alpha_{p}^{+}$ becomes shallower and
shallower and meanwhile more and more asymmetric as $\theta$
gradually reduces. Fig.~\ref{fig8}(b) shows that $\alpha_{p}^{-}$
changes in a way similar to $\alpha_{p}^{+}$ as far as their
spectral widths are concerned, \emph{i.e.} both become narrower as
$\theta$ gradually reduces. An evident change of $\alpha_{p}^{-}$
different from $\alpha_{p}^{+}$ lies in that the two-photon EIT dip
is absent for both $\theta=180^{\circ}$ and $\theta=160^{\circ}$ but
can be observed like $\alpha_{p}^{+}$ for $\theta=156.5^{\circ}$.
This can be understood by considering that $\theta=156.5^{\circ}$
corresponds to the case where both two-photon transitions
$|1\rangle\leftrightarrow|5\rangle$ and
$|2\rangle\leftrightarrow|5\rangle$ become Doppler free due to
$k_{p,c1}=k_{a,c2}^{\texttt{eff}}$ so that it is impossible to
attain the transmission non-reciprocity around four-photon resonance
without residual Doppler broadenings. As to the asymmetric features
of $\alpha_{p}^{+}$ and $\alpha_{p}^{-}$ spectra, they arise in fact
from effective Rabi frequencies $\Omega_{pe}^{\pm}$ and
$\Omega_{ce}^{\pm}$ as well as dynamic Stark shifts
$\Delta_{2d}^{\pm}$ and $\Delta_{5d}^{\pm}$, whose velocity
dependence cannot be eliminated via appropriate arrangements of the
probe, assistant, and coupling fields. Fortunately, the four
quantities exhibit quite small values and change just a little for
different atomic velocities so that the EIT dip remains well
developed. With above discussions, we conclude that the quenching of
$T_{p}^{+}$ for $\theta<157.3^{\circ}$ and the minimum of
$T_{p}^{-}$ at $\theta=157.3^{\circ}$ in Fig.~\ref{fig7} arise from
the Doppler-free asymmetric EIT spectra of $\alpha_{p}^{\pm}$ and a
slight velocity-dependent shift of the EIT dip away from four-photon
resonance.

\begin{figure}[ptbh]
\includegraphics[width=0.48\textwidth]{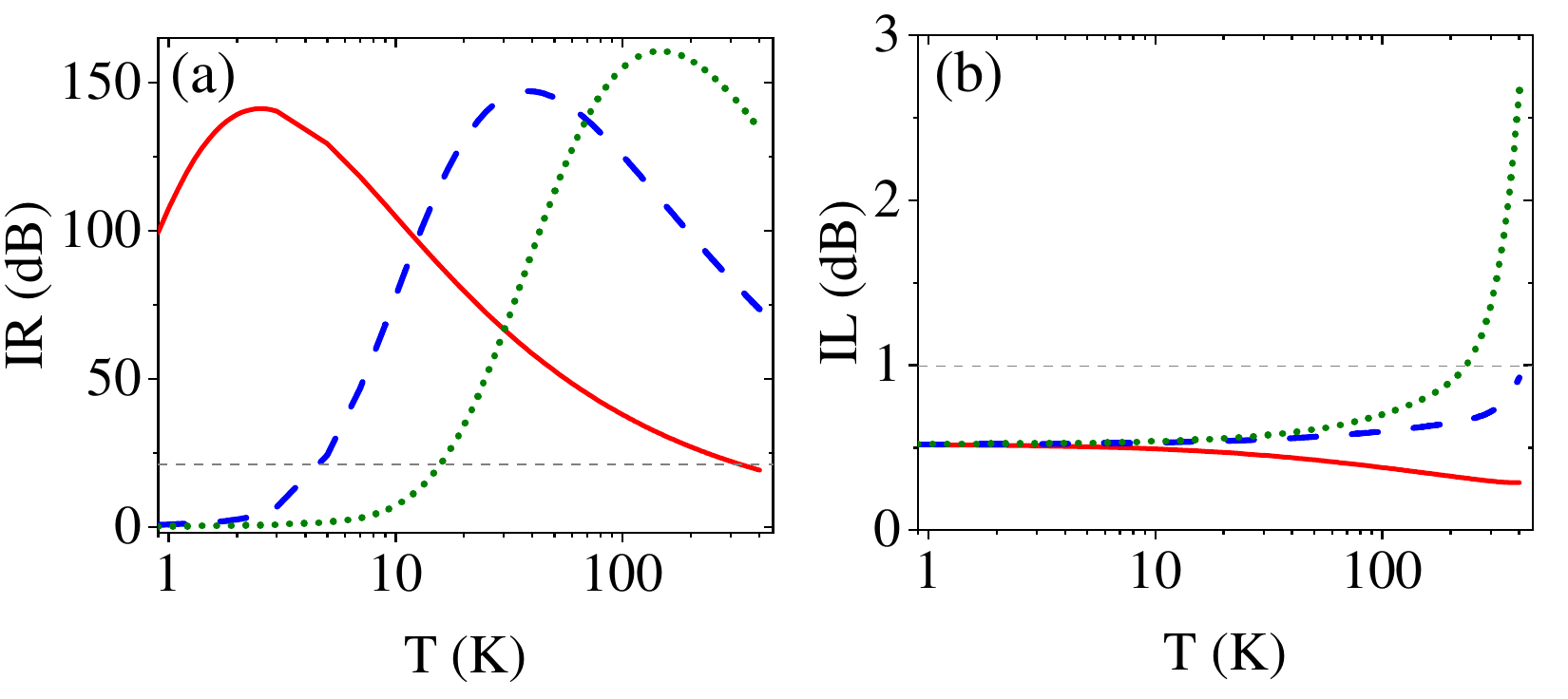}
\caption{\label{fig9} Isolation ratio $\texttt{IR}$ (a) and
insertion loss $\texttt{IL}$ (b) against temperature $T$ for
$\theta=180^{\circ}$ (red-solid), $160^{\circ}$ (blue-dashed), and
$158^{\circ}$ (green-dotted). Relevant parameters are the same as in
Fig.~\ref{fig3} except $\Delta_{p}=-1000$~MHz. Gray dotted lines
refer to $\texttt{IR}=20$ dB in (c) or $\texttt{IL}=1.0$ dB in (d)
as a reference.}
\end{figure}

What we observe in Fig.~\ref{fig7} and Fig.~\ref{fig8} answer
why the nonreciprocal bandwidth can be greatly enlarged by replacing
$\theta=180^{\circ}$ with $\theta=158^{\circ}$ at $T=300$ K as shown
in Fig.~\ref{fig5}. Finally, we examine in Fig.~\ref{fig9} the
joint effects of angle $\theta$ and temperature $T$ on the
transmission non-reciprocity in terms of isolation ratio
$\texttt{IR}$ and insertion loss $\texttt{IL}$ in the case of
$\Delta_{p}=-\Delta_{a}$. We can see that there is an optimal
temperature where $\texttt{IR}$ exhibits a maximum over $140$~dB for
each value of $\theta$ and this temperature is about $2.5$~K,
$35$~K, and $147$~K for $\theta=180^{\circ}$, $160^{\circ}$, and
$158^{\circ}$, respectively. As to $\texttt{IL}$, it monotonously
decreases for $\theta=180^{\circ}$ but continuously increases for
$\theta=160^{\circ}$ and $158^{\circ}$ as temperature $T$ becomes
larger. It is worth noting that we have $\texttt{IL}<1.0$~dB in a
wide range of temperature $T$ of our interest for
$\theta>160^{\circ}$ while $\texttt{IL}$ exceeds $1.0$~dB at
$T=235$~K for $\theta=158^{\circ}$. Anyway, this figure tells that
it is viable to attain a better trade-off between $\texttt{IR}$ and
$\texttt{IL}$ so as to realize a high-performance optical isolator
by simultaneously modulating angle $\theta$ and temperature $T$ in
appropriate ranges. The main benefit of such a joint modulation lies
in that it promises a flexible manipulation on both residual Doppler
broadenings of two-photon transitions and inevitable Doppler shifts
of effective Rabi frequencies and dynamic Stark shifts.

\section{\label{Conclusions}Conclusions}

In summary, we have investigated an efficient scheme for achieving
the magnet-free optical non-reciprocity in a three-level $\Lambda$
system dominated by two-photon transitions by considering the
free-space thermal $^{87}$Rb atoms as an example. The forward probe
field is found to experience a Doppler-free EIT window and hence
suffers very low losses in transmission, while the backward probe
field is strongly absorbed because the EIT window is smeared out as
the Doppler shifts on two-photon transitions $|1\rangle
\leftrightarrow |5\rangle$ and $|2\rangle \leftrightarrow |5\rangle$
don't cancel out again. It is of particular interest that the
non-reciprocal transmission may be well controlled by modulating the
frequency and amplitude of an assistant field as well as a common
misaligned angle between the two pairs of
$\{\omega_{p},\omega_{a}\}$ and $\{\omega_{c1},\omega_{c2}\}$
fields. It is also important that this transmission non-reciprocity
can exhibit high isolation ratios and low insertion losses in a wide
frequency range, benefiting from largely reduced Doppler broadenings
on two-photon transitions $|1\rangle \leftrightarrow |5\rangle$ and
$|2\rangle \leftrightarrow |5\rangle$. That means, our scheme allows
to manipulate hundreds of probe fields as multiple light
signals at the same time with similar isolation ratios and insertion
losses due to their insensitivities to single-photon detunings and
hence facilitate WDM applications in all-optical networks.

\section*{ACKNOWLEDGMENTS}
This work is supported by the National Natural Science Foundation of
China (Nos.~62375047 and 12074061).


\appendix
\section{\label{app} Absorption coefficient}
Here, we try to derive the two-photon absorption coefficient of a
probe field based on population $\rho_{55}$ in state $|5\rangle$ as
the five-level $\Lambda$ system in Fig.~\ref{fig1}(a) reduces to the
three-level $\Lambda$ system in Fig.~\ref{fig1}(b). To this end, we
first note that the probe field exhibits an intensity defined as
$I_{p}=c\epsilon_{0}E_{p}^{2}/2=2\hbar^{2}c
\epsilon_{0}|\Omega_{p}|^{2}/d_{13}^{2}$. Then, the number of
probe photons, passing through a section of the atomic sample at
position $z$, per time is given by
\begin{equation}\label{xid}
\frac{N_{p}(z)}{dt}=\frac{\pi
r_{p}^{2}}{\hbar\omega_{p}}I_{p}(z)=\frac{\hbar\epsilon_{0}\lambda_{p}
r_{p}^{2}}{d_{13}^{2}}|\Omega_{p}(z)|^{2},
\end{equation}
where $r_{p}$ denotes the probe beam radius. With this
consideration, we can further attain the number of photons lost per
time in an atomic slice from $z$ to $z+dz$
\begin{equation}\label{xid}
\frac{dN_{p}}{dt}=\frac{\hbar c\epsilon_{0}\lambda_{p}
r_{p}^{2}}{d_{13}^{2}}dI_{p},
\end{equation}
with $dN_{p}=N_{p}(z+dz)-N_{p}(z)$ and
$dI_{p}=|\Omega_{p}(z+dz)|^{2}-|\Omega_{p}(z)|^{2}$. Meanwhile, the
number of atoms in state $|5\rangle$ lost per time, due to
spontaneous decay after absorbing probe photons, in this atomic
slice is given by
\begin{equation}
\frac{\delta n_{a}}{dt}=N\pi r_{p}^{2}\rho_{55}\Gamma dz,
\end{equation}
where we have considered that the probe field interacts with $N\pi
r_{p}^{2}dz$ atoms in this slice of density $N$.

\begin{figure}[ptbh]
\includegraphics[width=0.48\textwidth]{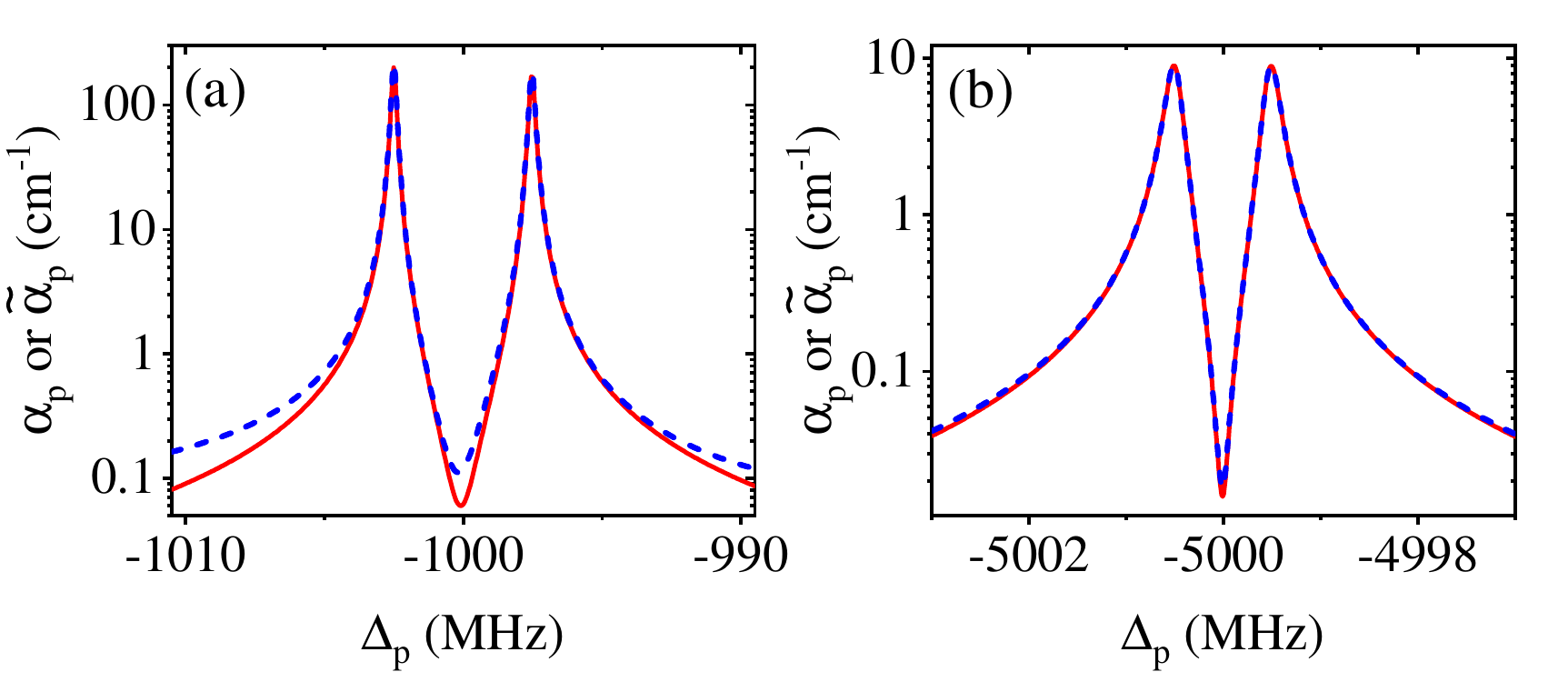}
\caption{\label{fig10} Absorption coefficients $\alpha_{p}$
(red-solid) and $\tilde{\alpha}_{p}$ (blue-dashed) as functions of
probe detuning $\Delta_{p}$ with $\Delta_{a}=\Delta_{c1}=1000$~MHz
(a) and $\Delta_{a}=\Delta_{c1}=5000$~MHz (b). Other parameters are
the same as in Fig.~\ref{fig2} except $\Omega_{a}=50$~MHz and
$\Delta_{c2}= -\Delta_{c1}-\Delta_{5d}+\Delta_{2d}$ changes with
$\Delta_{a}=\Delta_{c1}$.}
\end{figure}

According to the requirement of energy conservation
$dN_{p}/dt=-\delta n_{a}/dt$, we can derive the probe transmissivity
for an atomic sample of length $L$
\begin{equation}\label{transmission-cold}
T_{p}=\frac{I_{p}(L)}{I_{p}(0)}=\left|\frac{\Omega_{p}(L)}{\Omega_{p}(0)}\right|^{2}=e^{-\alpha_{p}L},
\end{equation}
with the absorption coefficient being
\begin{equation}\label{alpha-1}
\alpha_{p}=\frac{Nd_{13}^{2}}{\hbar\epsilon_{0}}\frac{\pi\Gamma}{\lambda_{p}}\frac{\rho_{55}}{|\Omega_{p}|^{2}},
\end{equation}
which should be $\Omega_{p}$-independent since
$\rho_{55}\propto|\Omega_{p}|^{2}$ in the limit of a weak probe
field. It is worth noting that the absorption coefficient may also
be expressed as
\begin{equation}\label{alpha-2}
\tilde{\alpha}_{p}=\frac{Nd_{13}^{2}}{\hbar\epsilon_{0}}\frac{2\pi}{\lambda_{p}}\frac{\mathrm{Im}
\rho_{31}}{\Omega_{p}},
\end{equation}
with $\rho_{31}\propto\Omega_{p}$ obtained by solving density
matrix equations of the original five-level $\Lambda$ system in the
steady state. Equations (\ref{alpha-1}) and (\ref{alpha-2}) allow us
to examine the validity of a reduced three-level $\Lambda$ system by
presenting a numerical comparison between the two absorption
coefficients with appropriate parameters as shown in
Fig.~\ref{fig10}. The results tell that absorption coefficients
$\alpha_{p}$ and $\tilde{\alpha}_{p}$ are in good agreement for
$|\Delta_{a,c1,c2}|/\Omega_{a,c1,c2}=20$ and fit better for
$|\Delta_{a,c1,c2}|/\Omega_{a,c1,c2}=100$ in the case of two-photon
near resonances $\Delta_{p}\simeq-\Delta_{a}$ and
$\Delta_{c1}\simeq-\Delta_{c2}$.

\end{document}